\documentclass[twocolumn]{aa}
\usepackage{graphicx}
\begin{document}
   \title{
H$_2$ absorption in a dense interstellar filament\\ in the Milky Way halo
}

   \author{P. Richter,
          \inst{1}
          K. R. Sembach,
          \inst{2}
          \and
          J. C. Howk
          \inst{3}
          }

   \offprints{P. Richter\\
   \email{richter@arcetri.astro.it}}

   \institute{
              Osservatorio Astrofisico di Arcetri, 
              Largo E. Fermi 5, 50125 Florence, Italy
         \and
              Space Telescope Science Institute, 
              3700 San Martin Drive, Baltimore, MD 21218, USA
         \and
              Center for Astrophysics \& Space Sciences, University
              of California, San Diego, 9500 Gilman Dr., 
              La Jolla, CA 92093, USA        
   }             

   \date{Received xxx; accepted xxx}

   \abstract{
   We investigate interstellar absorption 
   from molecular hydrogen (H$_2$) 
   and metals in an intermediate-velocity cloud 
   (IVC) in the direction of the LMC star 
   Sk -68\,80 (HD\,36521), based on data from the 
   {\it Far Ultraviolet Spectroscopic Explorer} (FUSE) satellite.
   H$_2$ absorption from the Lyman- and Werner bands is detected in 30
   lines at radial velocities $v_{\rm LSR} 
   \approx +50$ km\,s$^{-1}$ in this IVC
   that is presumably located in the Milky Way halo. 
   We obtain a total logarithmic
   H$_2$ column density of log $N($H$_2)=16.6 \pm 0.5$ 
   along with a very low Doppler parameter
   of $b=1.5^{+0.8}_{-0.2}$ km\,s$^{-1}$. 
   The presence of molecular material in this cloud is
   suprising, given the fact that the O\,{\sc i} column density
   (log $N($O\,{\sc i}$)=14.8 \pm 0.1$) implies 
   a very low neutral gas column density
   of $\sim 10^{18}$ cm$^{-2}$ (assuming a solar oxygen abundance). 
   If the H$_2$ column density represents its abundance in a 
   formation-dissociation equilibrium, the data imply
   that the molecular gas resides in a small, 
   dense filament at a volume density
   of $\sim 800$ cm$^{-3}$ and a thickness 
   of only 41 Astronomical Units (AU). The molecular 
   filament possibly
   corresponds to the tiny-scale atomic structures 
   (TSAS) in the diffuse interstellar medium
   observed in high-resolution optical data, H\,{\sc i}
   21cm absorption, and in CO emission.

   }
   \titlerunning{H$_2$ in a dense filament}
   \maketitle
%

\section{Introduction}

Recent high resolution absorption line measurements in the
optical and ultraviolet have shown that the interstellar
medium (ISM) consists of significant small-scale structure 
at sub-pc scales (e.g., Meyer \& Lauroesch 1999; Lauroesch
\& Meyer 1999; Lauroesch, Meyer, \& Blades 2000). Small-scale
structure can be identified as variations
in the shapes of absorption line profiles 
toward background sources at very small angular
separation (e.g., toward stellar clusters or binary stars)
and/or by re-observing the same background source at
different epochs. Optical depth variations on the scale
of several AUs are observed through H\,{\sc i} 21cm
absorption toward high-velocity pulsars and extragalactic
radio sources (Frail et al.\,1994; Faison et al.\,1998),
also indicating the presence of small-scale structure
in the ISM.
Such optical depth variations in local gas can introduce
significant systematic errors in observations for which 
foreground absorption has to be considered. For example,
reddening variations at arcmin scales are found to 
be responsible for substantial color variations among giant
branch stars in the Galactic globular cluster M22   
(Richter, Hilker, \& Richtler 1999); they are also observed
toward other globular clusters (e.g., von Braun et al.\,2002). 
All these observations
indicate that small-scale structure may represent an
important aspect of the ISM that yet is
poorly understood. 

\begin{figure*}[t!]
\resizebox{1.00\hsize}{!}{\includegraphics{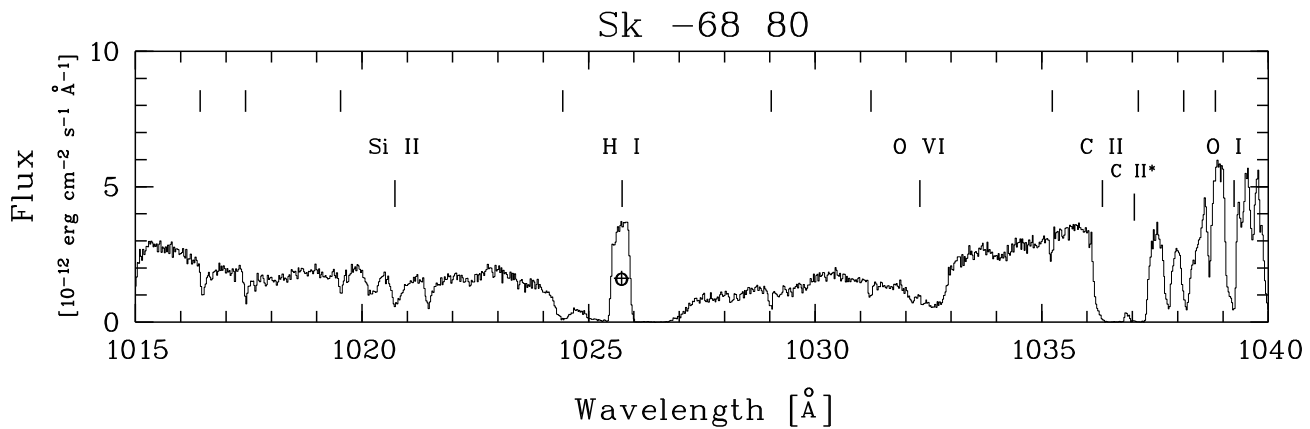}}
\caption[]{
A portion of the FUSE spectrum of Sk -68\,80 in the
wavelength range between 1015 and 1040 \AA. Interstellar
atomic absorption lines are labeled above the spectrum.
The unlabeled tic marks indicate interstellar H$_2$ absorption
lines from the Lyman and Werner bands.
}
\end{figure*}

Optical and ultraviolet absorption spectroscopy of
stars and extragalactic background sources is a very 
sensitive method to study small-scale structure in the
diffuse ISM because of the large number of spectral lines
that are available for this task, sampling the molecular,
neutral and ionized gas phases at a very high accuracy.
Particulary interesting for this purpose are observations 
of diffuse halo clouds that typically have H\,{\sc i}
column densities $\leq 5 \times 10^{20}$ cm$^{-2}$ 
(see Wakker 2001) and which are
well separated in radial-velocity ($|v_{\rm LSR}|>30$ km\,s$^{-1}$)
from the strong local disk absorption.
These intermediate- and high-velocity clouds 
(IVCs and HVCs, respectively) trace 
various processes that circulate gas through the Milky
Way halo, such as the ``Galactic Fountain'' (Shapiro \&
Field 1976; Houck \& Bregman 1991), infall of low-metallicity
gas from intergalactic space (Wakker et al.\,1999; Richter 
et al.\,2001a, 2001b), and interaction with the Magellanic Clouds
(Lu et al.\,1998). Spanning a wide range of metallicities 
and not being exposed to intense stellar UV radiation, IVCs
and HVCs serve as important interstellar laboratories to
study physical processes in the diffuse ISM. 

In this paper we use far-ultraviolet (FUV) absorption
line data from the {\it Far Ultraviolet Spectroscopic
Explorer} (FUSE) obtained to study molecular and atomic
absorption in intermediate- and high-velocity gas toward the
Wolf-Rayet star Sk -68\,80 (HD\,36521) in the Large Magellanic 
Cloud (LMC). We emphasize molecular hydrogen (H$_2$) 
absorption in the IVC gas that possibly samples small-scale
structure at sub-pc levels (Richter et al.\,2003). 
This paper is organized as follows: 
In Sect.\,2 we describe the FUSE observations,
the data reduction, and the analysis method.  
In Sect.\,3 we review the sight-line structure in the direction of
Sk -68\,80. In Sect.\,4 we analyze in detail absorption by metals and H$_2$ 
in the IVC component at $v_{\rm LSR} \approx +50$ km\,s$^{-1}$. 
In Sect.\,5 we briefly consider atomic and molecular absorption in the HVC component 
near $+120$ km\,s$^{-1}$. 
A discussion of 
our results is presented in Sect.\,6.
In the Appendix we review other
LMC sight lines; in particular we re-analyze the
FUV spectrum of Sk -68\,82 (HD\,269546).

\section{Observations and data handling}

The LMC star Sk -68\,80 (HD\,36521; $l=279.33, b=-32.79$)
is a Wolf-Rayet star (WC4+O6.5; $V=12.40$), 
and part of the OB association LH\,58 in the 
H\,{\sc ii} region N\,144 
(Massey, Waterhouse, \& DeGioia-Eastwood 2000).
FUSE observations of Sk -68\,80 (program ID P1031402)
were carried out on 17 December 1999 through the 
large aperture (LWRS) of the FUSE instrument. Four 
exposures were taken, totalling 9.7 ks of integration
time. FUSE is equipped with four co-aligned Rowland circle
spectrographs
and two microchannel-plate detectors,
covering the wavelength region between 905 and 1187 \AA.
Two of the four available channels are coated with Al+LiF (for maximum
throughput at $\lambda >1000$ \AA), the other two with
SiC (for $\lambda \leq 1000$ \AA). The LiF and SiC channels
and their segments overlap in the
wavelength region between $1000$ and $1100$ \AA.
FUSE provides three entrance apertures:
$30\farcs0 \times 30\farcs0$ (LWRS; the one used for 
the observations presented here),      
$4\farcs0 \times 20\farcs0$ (MDRS), and  
$1\farcs25 \times 20\farcs0$ (HIRS). 
Details about the instrument and its on-orbit performance 
are presented by Moos et al.\,(2000) and Sahnow et al.\,(2000).
The FUSE spectrum of Sk -68\,80 was recorded in photon-address
mode (storing the arrival time, the pulse height and the 
X/Y location of each detection) and was reduced with the
v2.0.5 version of the {\tt CALFUSE}
standard pipeline, which corrects for the detector
background, orbital motions of the spacecraft, and geometrical
distortions. We find that the wavelength calibration provided
by {\tt CALFUSE} v2.0.5, as well as the spectral resolution (as
checked by fitting Gaussian absorption profiles to the data),
has significantly improved the data quality for Sk -68\,80
in comparison to reductions with earlier {\tt CALFUSE} versions.
For the purpose of this study (measuring weak absorption 
components in a multi-component absorption pattern; see
next section), this improvement is crucial in view
of the determination of precise equivalent widths and the
separation of the various absorption components.
The reduced FUSE data of Sk -68\,80 have a velocity resolution
of $\sim 20$ km\,s$^{-1}$ (FWHM), corresponding to a
resolving power of $R=\lambda / \Delta \lambda \approx 15,000$.
Radial velocities were transformed into the Local Standard
of Rest (LSR) system. For this spectrum, we estimate an uncertainty of $\sim 10$
km\,s$^{-1}$ ($1\sigma$) for the velocity 
calibration provided by the {\tt CALFUSE}
v2.0.5 pipeline.

The average continuum flux is
$\sim 2 \times 10^{-12}$ erg\,cm$^{-2}$\,s$^{-1}$\,\AA$^{-1}$,
resulting in a signal-to-noise ratio (S/N)
of $\sim 20$ per resolution element. The individual exposures
were co-added, and the data were rebinned to
$6$ km\,s$^{-1}$ wide bins (3 pixel rebinning). 
Fig.\,1 shows the FUSE spectrum of Sk -68\,80 in the 
wavelength range between $1015$ and $1040$ \AA.
Atomic absorption features are labeled above the spectrum;
molecular hydrogen absorption lines are marked with
tic marks above the metal line identifications. 
The continuum flux of Sk -68\,80                  
varies on large scales  ($>5$ \AA) in the   
FUSE spectrum. On smaller scales ($<1$ \AA) 
the continuum is relatively smooth, making
the continuum placement for interstellar absorption
towards Sk -68\,80 relatively reliable, in contrast to many other 
LMC stars that have strongly varying continua even
in the sub-\AA\,regime (e.g., Sk -68\,82, see Appendix).
The continuum was fitted locally for each measured
absorption line, using low-order polynomials. 
Equivalent widths of the absorption components 
were measured by fitting multi-component Gaussian profiles
to the data. Column densities were derived using a 
standard curve-of-growth technique.

\section{The sight-line structure towards Sk -68\,80}

Although the LMC provides an excellent set of
stellar backgound sources for the study of 
intermediate- and high-velocity halo gas, 
the complex sight-line structure toward the
LMC makes the analysis of foreground halo
components a difficult task (e.g., Savage \& de\,Boer 1979). 

Fig.\,2 shows FUSE absorption profiles of
O\,{\sc i} $\lambda 1039.230$, 
Fe\,{\sc ii} $\lambda 1144.989$,
P\,{\sc ii} $\lambda 1152.818$,
and C\,{\sc i} $\lambda 945.191$
in the direction of Sk -68\,80, plotted
on the LSR
velocity scale.
The various absorption components can be divided into three
different groups: (1) absorption by local Milky Way gas
in the velocity range between $v_{\rm LSR}
=-40$ to $+40$ km\,s$^{-1}$, (2) absorption by
intermediate- and high-velocity clouds
at velocities near $+50$ to $+60$ km\,s$^{-1}$ (IVC) and
$+90$ to $+140$ km\,s$^{-1}$ (HVC), and (3) absorption by LMC gas
in the velocity range between
$+160$ to $+350$ km\,s$^{-1}$, with a component structure 
that generally varies over the field of the 
LMC, as seen toward other background sources (Tumlinson et
al.\,2002; Howk et al.\, 2002). While the IVC gas
in front of the LMC most likely belongs to the Milky Way,
the origin of the HVC is not clear. 
This cloud could be Galactic fountain gas 
(Richter et al.\,1999; Welty et al.\,1999)
or high-velocity gas that has been pushed out 
of the LMC (Staveley-Smith et al.\,2003).

The individual absorption components exhibit
sub-structure, which is clearly visible in Fig.\,2 in the
Fe\,{\sc ii} absorption of the local Milky Way 
component and the LMC component. It is very likely
that most of the existing sub-components are not
resolved in the FUSE data, a fact that has to be taken
into account for the interpretation of the observed
absorption pattern. Indeed, Welty et al.\,(1999) find
at least 46 absorption components in the direction of the 
LMC SN1987A using high-resolution (FWHM$<3$ km\,s$^{-1}$)
optical spectra, which emphasizes the extreme complexity 
of the sight-line structure in the general direction of
the LMC.  

In this paper, we concentrate on absorption from the IVC and HVC
gas. At the resolution of FUSE, both halo components are
well fitted by single Gaussian profiles, with
central velocities near $+50$ km\,s$^{-1}$ (IVC)
and $+120$ km\,s$^{-1}$ (HVC), as marked in Fig.\,2 
with dotted lines. 
Because of the large number of absorption features
in the spectrum, severe blending problems
occur for many atomic and molecular lines; thus, possible 
blending effects have to be considered carefully for 
each absorption line. H\,{\sc i} 
21cm data for the IVC and HVC material towards 
N\,144 is available from a Parkes spectrum (32 arcmin beam) centered
on Sk -68\,82, $\sim 1$ arcmin away from 
Sk -68\,80 (McGee \& Newton 1986). The Parkes data show
the IVC component at an H\,{\sc i} column density
of $N($H\,{\sc i}$)_{\rm IVC}\approx 4 \times 10^{18}$ cm$^{-2}$,
while the HVC component has 
$N($H\,{\sc i}$)_{\rm HVC}\approx 1 \times 10^{19}$ cm$^{-2}$
(McGee \& Newton 1986). Newer Parkes data 
($\sim 15$ arcmin beam), however,
imply lower column densities of
$N($H\,{\sc i}$)_{\rm IVC}\leq 2 \times 10^{18}$ cm$^{-2}$
and
$N($H\,{\sc i}$)_{\rm HVC}\approx 9 \times 10^{18}$ cm$^{-2}$
in the direction of Sk -68\,80/Sk -68\,82.
The differences in column densities may indicate 
the existence of H\,{\sc i} sub-structure
on scales between 15 and 32 arcmin 
($\sim 9$ pc at a distance of 2 kpc).
Therefore, the 21cm data most likely provide 
only a rough estimate of the H\,{\sc i} column densities
in the IVC and HVC
towards Sk -68\,80. The
H\,{\sc i} radio data suggest, however, that
the column densities of the neutral gas within
the two halo clouds in front of the LMC
are relatively low when compared
to other Galactic IVC and HVC complexes
(see Wakker 2001).

\begin{figure}[t!]
\resizebox{1.0\hsize}{!}{\includegraphics{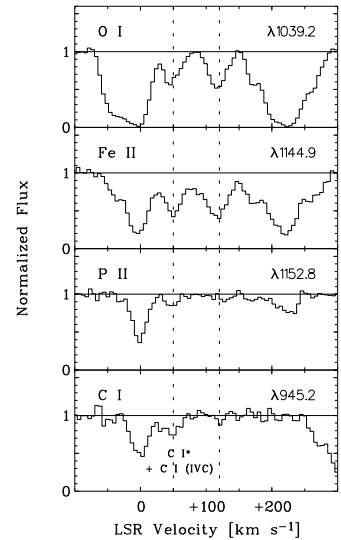}}
\caption[]{
Atomic absorption profiles from O\,{\sc i}, Fe\,{\sc ii},
P\,{\sc ii}, and C\,{\sc i} in the direction of
Sk -68\,80, plotted against LSR velocity. 
Galactic IVC and HVC absorption occurs
near $+50$ km\,s$^{-1}$ and $+120$ km\,s$^{-1}$
(dotted lines),
respectively. Local disk gas is seen at velocities
$<40$ km\,s$^{-1}$; absorption from LMC gas 
occurs at velocities $>160$ km\,s$^{-1}$.
Blending lines are labeled within the plot.
}
\end{figure}
               
\section{The IVC at near $+50$ km\,s$^{-1}$}

\subsection{Metal absorption}

We have measured equivalent widths for IVC absorption
in 13 lines of C\,{\sc i}, 
O\,{\sc i}, Si\,{\sc ii}, P\,{\sc ii},
Ar\,{\sc i}, and Fe\,{\sc ii}, as listed in Table 1.
The six Fe\,{\sc ii} lines that are detected in the
IVC component fit on a single-component 
curve of growth with a Doppler parameter
$b=5.9^{+2.8}_{-1.3}$ km\,s$^{-1}$ and a column
density of log $N($Fe\,{\sc ii}$)=14.1 \pm 0.1$ (Fig.\,3).
Fitting the lines of O\,{\sc i}, Si\,{\sc ii}, P\,{\sc ii},
and Ar\,{\sc i} (see Table 1) to the same curve of growth,
we derive log $N$(O\,{\sc i})$=14.8\pm0.1$,
log $N$(Si\,{\sc ii})$=14.3\pm0.2$, log $N$(P\,{\sc ii})$=12.8\pm0.1$,
and log $N$(Ar\,{\sc i})$=12.9\pm0.1$. 
Column densities are also listed in Table 2.
Unfortunately, C\,{\sc i}
$\lambda 945.191$ absorption at IVC velocities is blended by
Galactic C\,{\sc i}$^{\star}$ (see Fig.\,2);
however, assuming that all of the absorption is due to 
C\,{\sc i} in the IVC, we obtain an upper limit of 
log $N$(C\,{\sc i})$\leq13.4$.
It is
possible that the $b$ values for these species are slightly different
than that of Fe\,{\sc ii}, depending on the ionization
structure of the cloud. However, due to the limited number of
lines and the lack of further information we have to adopt
$b=5.9$ km\,s$^{-1}$ for all atomic species.

The ratios of [Fe\,{\sc ii}/O\,{\sc i}]$=+0.7$ and
[Si\,{\sc ii}/O\,{\sc i}]$=+0.8$
(where [Fe/O]$=0.0$ represents the
solar abundance ratio on a logarithmic scale;
Anders \& Grevesse 1989; Grevesse \& Noels 1993) 
are relatively high, indicating a substantial amount
of ionized gas that is sampled by Fe\,{\sc ii}
and Si\,{\sc ii} (ionization potentials are 16.2 and
16.4 eV, respectively), but not by O\,{\sc i} (ionization
potential is 13.6 eV, identical with that of H\,{\sc i}).
The data for Fe\,{\sc ii} and Si\,{\sc ii} suggest that the
column density of ionized gas in the IVC exceeds that of the neutral
gas by a factor of $\sim 6$. This factor could be even higher if
some of the Fe and Si is depleted onto dust grains.
This high degree of ionization may indicate the presence
of shocks.
Due to the uncertainty of the H\,{\sc i} column density in the
IVC towards Sk -68\,80 from the 21cm data (see previous section) 
and the high column density of ionized gas we refrain 
from calculating gas phase abundances for these elements.
Instead, we use the O\,{\sc i} absorption line data to
estimate the H\,{\sc i} column
density in the IVC along this sight-line, 
which will be important for the interpretation
of the H$_2$ abundance in the IVC (next section). 
The abundances of O and H are
coupled by charge exchange reactions. Moreover, 
oxygen does not significantly deplete onto dust grains.
Assuming an intrinsic oxygen abundance, we can 
use $N$(O\,{\sc i}) to obtain the H\,{\sc i} column density
in the IVC.
Previous studies of Galactic IVCs
(Richter et al.\,2001a, 2001c) indicate that these
clouds have solar metal abundances, 
suggesting that they originate
in the disk of the Milky Way.
If we assume that the IVC in front of the LMC
also has a solar oxygen abundance, and take
log (O/H)$_{\rm solar}=-3.26$ (Holweger 2001),
we derive 
$N($H\,{\sc i}$) \approx 1 \times 10^{18}$ cm$^{-2}$.
This value is compatible with the upper limit for 
$N($H\,{\sc i}) from the newer Parkes 
data (see Sect.\,3).

\begin{table}[t!]
\caption[]{Equivalent widths$^{a}$ for IVC and HVC absorption towards Sk -68\,80}
\begin{tabular}{lrrll}
\hline
\hline
Species     & $\lambda_{\rm vac}$\,$^{\rm b}$ & log\,$\lambda f^{\rm b}$ & 
$W_{\lambda}$\,$_{\rm IVC}$ & $W_{\lambda}$\,$_{\rm HVC}$ \\
& [\AA] &  & [m\AA] & [m\AA]\\
\hline
C\,{\sc i}            &  945.191 & 2.411 & $\leq23$  & $\leq10$  \\
O\,{\sc i}            &  948.686 & 0.778 & $21\pm5$  & $31\pm6$ \\
                      & 1039.230 & 0.980 & $37\pm5$  & $49\pm4$  \\
Si\,{\sc ii}          & 1020.699 & 1.225 & $23\pm7$ & $9\pm4$ \\                     
P\,{\sc ii}           & 1152.818 & 2.451 & $14\pm3$  & $\leq6$ \\
Ar\,{\sc i}           & 1048.220 & 2.440 & $15\pm3$ & $\leq 10$ \\
Fe\,{\sc ii}          & 1055.262 & 0.962 & $10\pm3$  & $5\pm3$  \\
                      & 1096.877 & 1.554 & $28\pm5$  & $28\pm5$  \\
                      & 1121.975 & 1.512 & $24\pm5$  & ...   \\
                      & 1125.448 & 1.244 & $16\pm3$  & $15\pm4$  \\
                      & 1142.366 & 0.633 & $4\pm2$  & $4\pm2$   \\
                      & 1143.226 & 1.342 & $26\pm3$  & $12\pm3$  \\
                      & 1144.938 & 2.096 & $59\pm4$ & $104\pm5$  \\
\hline
H$_2$ R(0),4-0 & 1049.366 & 1.383 & $50\pm12$   & ... \\
H$_2$ R(0),1-0 & 1092.194 & 0.814 & $39\pm7$   & ... \\
H$_2$ R(0),0-0 & 1108.128 & 0.283 & $26\pm8$   & ... \\
\hline
H$_2$ R(1),8-0 & 1002.457 & 1.256 & $27\pm10$  & $\leq 19$ \\  
H$_2$ P(1),8-0 & 1003.302 & 0.948 & $33\pm8$   & ... \\
H$_2$ R(1),4-0 & 1049.960 & 1.225 & $32\pm5$   & $\leq 14$ \\
H$_2$ P(1),4-0 & 1051.033 & 0.902 & $20\pm4$   & ... \\
H$_2$ P(1),3-0 & 1064.606 & 0.805 & $29\pm6$\,$^{\rm c}$ & ... \\
H$_2$ R(1),2-0 & 1077.702 & 0.919 & $30\pm7$  & ...\\
H$_2$ R(1),1-0 & 1092.737 & 0.618 & $25\pm6$  & ... \\
H$_2$ R(1),0-0 & 1108.639 & 0.086 & $26\pm8$  & ... \\
\hline
H$_2$ R(2),8-0 & 1003.989 & 1.232 & $14\pm3$ & $\leq 11$ \\
H$_2$ P(2),8-0 & 1005.398 & 0.993 & $17\pm5$ & ... \\
H$_2$ Q(2),0-0 & 1010.938 & 1.385 & $16\pm5$ & $\leq 10$ \\
H$_2$ R(2),7-0 & 1014.980 & 1.285 & $14\pm4$ & ... \\
H$_2$ P(2),7-0 & 1016.466 & 1.007 & $28\pm15$\,$^{\rm c,d,f}$ & ... \\
H$_2$ R(2),4-0 & 1051.498 & 1.168 & $12\pm3$ & ... \\
H$_2$ P(2),4-0 & 1053.284 & 0.982 & $13\pm4$ & ... \\
H$_2$ P(2),2-0 & 1081.269 & 0.708 & $11\pm4$\,$^{\rm f}$ & ... \\
\hline
H$_2$ R(3),0-0 & 1010.128 & 1.151 & $12\pm3$ &  $\leq 9$ \\
H$_2$ R(3),7-0 & 1017.427 & 1.263 & $22\pm8$ & $16\pm4$\,$^{\rm e}$ \\
H$_2$ P(3),6-0 & 1031.195 & 1.055 & $14\pm4$ &  $11\pm4$ \\
H$_2$ R(3),5-0 & 1041.159 & 1.222 & $15\pm3$ & ... \\
H$_2$ P(3),5-0 & 1043.504 & 1.060 & $14\pm3$ & ... \\
H$_2$ R(3),4-0 & 1053.975 & 1.137 & $16\pm4$ & ... \\
H$_2$ P(3),4-0 & 1056.471 & 1.006 & $12\pm3$ & ... \\
H$_2$ P(3),3-0 & 1070.141 & 0.910 & $15\pm3$ & $\leq 8$\\
H$_2$ P(3),1-0 & 1099.792 & 0.439 & $10\pm4$ & $7\pm3$ \\
\hline
H$_2$ R(4),5-0 & 1044.543 & 1.195 & $6\pm3$ & $\leq 9$ \\
H$_2$ R(4),4-0 & 1057.379 & 1.138 & $5\pm2$ & $\leq 8$\\
\hline
\end{tabular}
\noindent

$^{\rm a}$ Equivalent widths, $1\sigma$ errors and $3\sigma$ 
upper limits are given.\\
$^{\rm b}$ Wavelengths and oscillator strengths from 
Morton (1991), Morton (2003, in preparation), and Abgrall \& Roueff (1989).
$^{\rm c}$ Suprisingly strong.
$^{\rm d}$ H$_2$ absorption extends to $+100$ km\,s$^{-1}$.
$^{\rm e}$ Possibly blended by H$_2$ W Q(5),0-0.
$^{\rm f}$ Line not included in curve-of-growth fit.
\end{table}

\subsection{H$_2$ absorption}

Molecular hydrogen absorption in the IVC component at
$+50$ km\,s$^{-1}$ is found in 30 transitions in
the Lyman- and Werner electronic bands. 
IVC H$_2$ absorption is present in rotational levels
$J=0-4$ at relatively low equivalent widths ($W_{\lambda}
\leq 70$ m\AA). The fact that weak H$_2$ absorption 
occurs in so many lines that span a wide range 
in oscillator strengths (see Abgrall \& Roueff 1989) already
indicates that most lines must lie on the flat part
of a curve of growth with a very low $b$ value. 
Table 1 presents equivalent widths of 30
H$_2$ lines from IVC H$_2$ absorption above a 
$2\sigma$ detection level. 
Fig.\,4 shows a selection of H$_2$ absorption
profiles plotted on the LSR velocity scale.
The data points fit best
on a curve of growth with logarithmic H$_2$ column
densities, log $N(J)$, of log $N(0)=16.5^{+0.1}_{-0.5}$,
log $N(1)=16.0^{+0.3}_{-0.5}$, log $N(2)=14.4^{+0.3}_{-0.2}$,
log $N(3)=14.6^{+0.3}_{-0.2}$, and log $N(4)=13.7^{+0.1}_{-0.1}$,
and a $b$ value of
$1.5^{+0.8}_{-0.2}$ km\,s$^{-1}$ (see also Table 2). Such a low
$b$ value is unusual for Galactic halo clouds (e.g., 
Richter et al.\,2003), implying that the H$_2$ gas
is located in a relatively confined region with little 
interstellar turbulence.
In order to check the quality of the fit
we also have
put the data on curves of growth with higher
$b$ values. The H$_2$ data points, however, are confined in a relatively
narrow region in the log ($W_{\lambda}/\lambda)-$log ($Nf\lambda$)
parameter space, and all fits to curves of growth with $b>1.5$ km\,s$^{-1}$
lead to very unsatisfying results. We thus 
adopt $b=1.5$ km\,s$^{-1}$
for the discussion below. Adding the individual column
densities for $N(J)$ given above, the total H$_2$ column density
is log $N($H$_2)=16.6\pm0.5$, the highest found for IVC gas so far
(see Richter et al.\,2003). 
The fraction of hydrogen in molecular form, 
$f=2N($H$_2)/(N$(H\,{\sc i}$)+2N($H$_2))$,
can be estimated only indirectly, 
since the H\,{\sc i} column density from the
21cm observations probably does not give 
a reliable estimate for $N$(H\,{\sc i}$)$
in the IVC towards Sk -68\,80 (see Sect.\,4.1).
If we assume that
$N($H\,{\sc i}$)=10^{18}$ cm$^{-2}$, as estimated
from O\,{\sc i} in the previous section, 
we obtain 
$f\approx 0.07$.
Thus, in comparison to previous IVC H$_2$ results,
the molecular hydrogen fraction is remarkably high,
especially in light of the fact that the neutral
gas column density appears to be rather small, and
that most of the IVC gas is ionized.

\begin{figure*}[t!]
\resizebox{1.0\hsize}{!}{\includegraphics{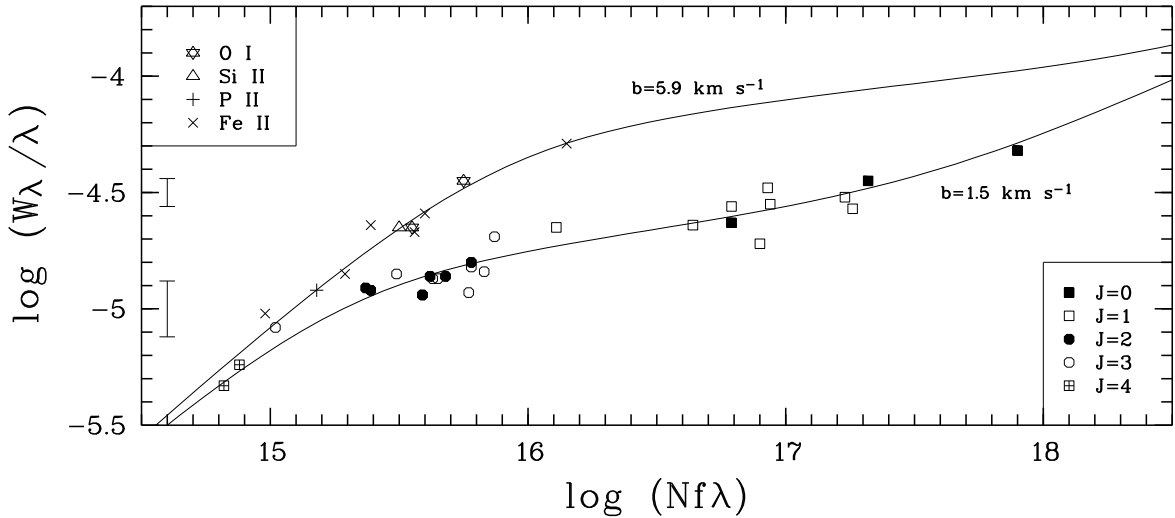}}
\caption[]{
Curves of growth (COG) for atomic and molecular absorption in the IVC
towards Sk -68\,80. Atomic absorption lines (identified
in the upper left corner) fit on a curve of growth with
$b=5.9^{+2.8}_{-1.2}$ km\,s$^{-1}$. Molecular hydrogen
lines (identified in the lower right corner) fit on
a COG with a much lower $b$ value of 
$1.5^{+0.8}_{-0.2}$ km\,s$^{-1}$. Typical error bars for
the data points in the different regions of
the COG are shown on the left hand side.
}
\end{figure*}

We now analyze
the rotational excitation of the H$_2$ gas. In Fig.\,5 we have
plotted the H$_2$ column density for each rotation level, $N(J)$,
divided by the quantum mechanical statistical weight, $g_J$, 
against the rotational excitation energy, $E_J$.
The data points follow the usual trend that is seen for many H$_2$ 
absorption line measurements: the two rotational ground states
($J=0$ and $1$) lie on a straight line that represents the
Boltzmann distribution for a temperature of T$_{\rm 01}$, whereas
a different Boltzmann fit with T$_{\rm 24}>$T$_{\rm 01}$ is required
to describe the level population for $J=2-4$. We
obtain T$_{\rm 01}=51\pm11$ K and T$_{\rm 24}=532\pm124$ K.
The value for T$_{\rm 01}$ probably reflects the kinetic temperature
of the H$_2$ gas, implying that H$_2$ line self-shielding 
is protecting the interior of the cloud from being
excited and dissociated by UV photons. The value of 
$51$ K is lower than found {\it on average} in local disk gas
($\sim 80$ K) and on the 
lower side of the distribution of kinetic temperatures
in local diffuse H$_2$ gas (Savage et al.\,1977). UV photon pumping and 
H$_2$ formation pumping (see, e.g., Shull \& Beckwith 1982)
are believed to excite the higher rotational states
($J\geq 2$) of the H$_2$, resulting in an equivalent Boltzmann
temperature (532 K) that is much higher than that for the 
rotational ground states. 
In view of the relatively mild UV radiation field 
in the halo (see discussion in Sect.\,4.3), the enhanced excitation most likely is 
caused by the formation process of H$_2$ on the surface
of dust grains, although other processes, such as shocks,
may also play a role here.

\begin{table}[t!]
\caption[]{Column densities$^{\rm a}$  and $b$ values for the IVC and HVC towards Sk -68\,80}
\begin{tabular}{lrrrr}
\hline
\hline
Species & log $N_{\rm IVC}$ & $b_{\rm IVC}$ & log $N_{\rm HVC}$ & $b_{\rm HVC}$ \\
        &                   & [km\,s$^{-1}$] &                 & [km\,s$^{-1}$] \\
\hline
C\,{\sc i}   &  $\leq 13.4$ & $5.9^{+2.9}_{-1.3}$ & $\leq 12.4$ & $30.0^{+9.2}_{-5.4}$ \\ 
O\,{\sc i}   &  $14.8\pm0.1$ &                    & $14.8\pm0.1$ & \\
Si\,{\sc ii} &  $14.3\pm0.2$ &                    & $13.8\pm0.1$ & \\
P\,{\sc ii}  &  $12.8\pm0.1$ &                    & $\leq 12.3$ & \\
Ar\,{\sc i}  &  $12.9\pm0.1$ &                    & $\leq 12.6$ & \\
Fe\,{\sc ii} &  $14.1\pm0.1$ &                    & $14.0\pm0.1$ & \\
\hline
H$_2$\,$J=0$ & $16.5^{+0.1}_{-0.5}$ & $1.5^{+0.8}_{-0.2}$ & $\leq 15.6$ & $\geq 1$ \\
H$_2$\,$J=1$ & $16.0^{+0.3}_{-0.5}$ &                     & $\leq 15.1$ & \\
H$_2$\,$J=2$ & $14.4^{+0.3}_{-0.2}$ &                     & $\leq 14.2$ & \\
H$_2$\,$J=3$ & $14.6^{+0.3}_{-0.2}$ &                     & $\leq 15.3$ & \\
H$_2$\,$J=4$ & $13.7^{+0.1}_{-0.1}$ &                     & $\leq 14.2$ & \\
H$_2$\,total & $16.6^{+0.5}_{-0.5}$ &                     & $\leq 15.6$ & \\
\hline
\end{tabular}
\noindent

$^{\rm a}$ Column densities, $1\sigma$ errors and $3\sigma$
upper limits are given
\end{table}

\begin{figure*}[t!]
\resizebox{1.0\hsize}{!}{\includegraphics{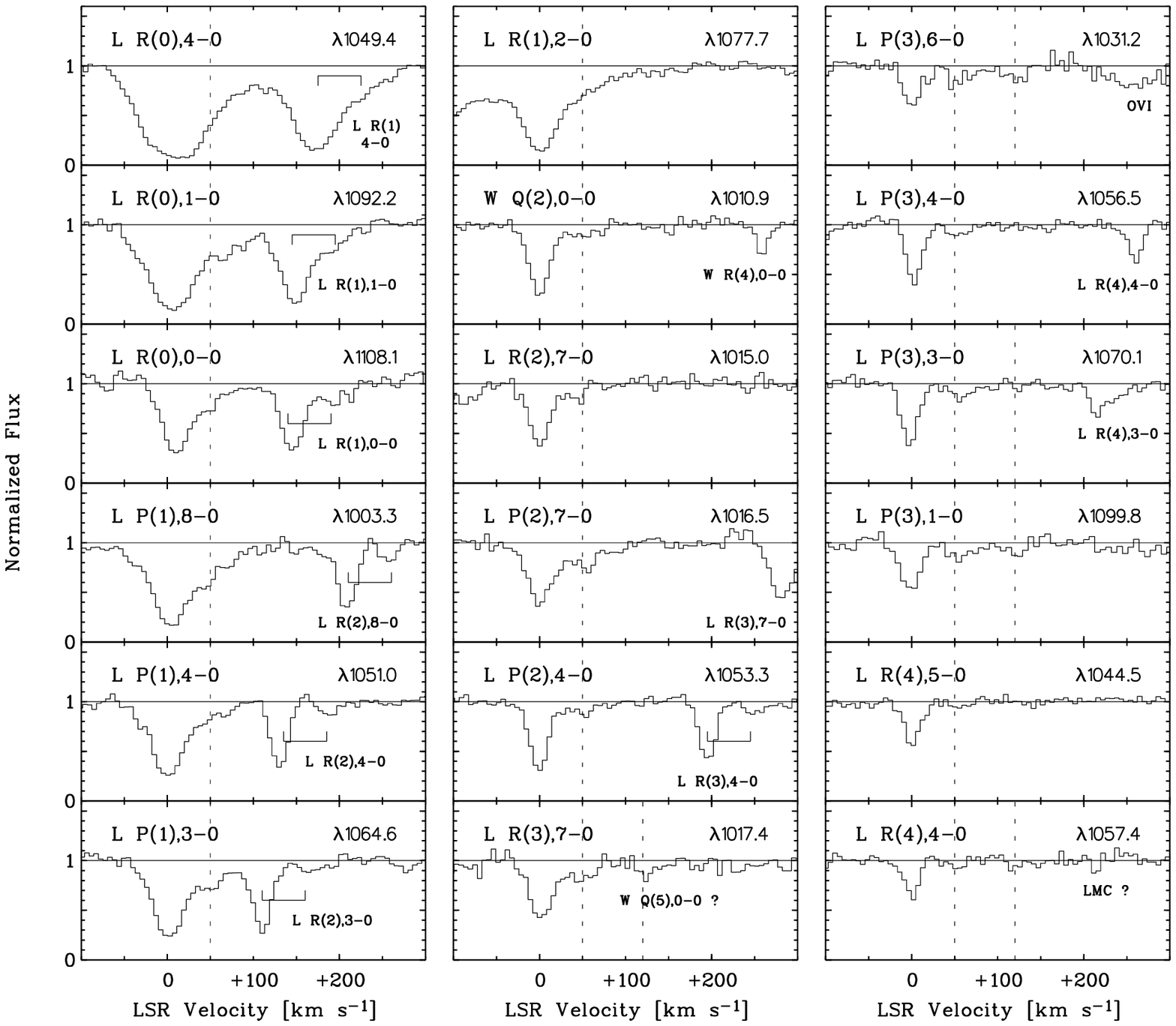}}
\caption[]{
Selection of normalized H$_2$ absorption profiles towards Sk -68\,80. 
The dotted lines indicate H$_2$ absorption by
intermediate-velocity halo gas near $+50$ km\,s$^{-1}$
and possibly high-velocity halo gas near $+120$ km\,s$^{-1}$
(5 cases). H$_2$ absorption from local Galactic gas occurs near
$0$ km\,s$^{-1}$.
Blends of other lines and their components are also
labeled. 
}
\end{figure*}

\subsection{Physical properties of the H$_2$ gas}

H$_2$ absorption in intermediate-velocity halo gas is a widespread
phenomenon, as is shown in the FUSE survey of molecular hydrogen
in H\,{\sc i} IVCs (Richter et al.\,2003,
hereafter R03). The IVC H$_2$ survey data suggest that
the possibility of intersecting intermediate-velocity
H\,{\sc i} gas containing molecular material may be as 
high as 50 percent. The findings so far imply
a very diffuse molecular gas phase with molecular hydrogen fractions
typically below $f=10^{-3}$ (R03).

A simple model for the H$_2$ abundance in IVCs in a formation-dissociation
equilibrium (R03) requires that the H$_2$ resides in small ($\sim 0.1$ pc),
dense ($\sim 30$ cm$^{-3}$) gas blobs or filaments.
In comparison to the
previous IVC H$_2$ detections, the present measurement in the IVC towards Sk -68\,80 
stands out because the H\,{\sc i} column density of this component
appears to be exceptionally low ($N$(H\,{\sc i})$\approx 10^{18}$ cm$^{-2}$),
as indicated by the low O\,{\sc i} column density and the 21cm
data (see Sect.\,3). 
Below we suggest 
an answer to the question of how 
molecular gas can form in such a low column density environment
and can avoid the UV photo dissociation. 
We assume that the $+50$ km\,s$^{-1}$ absorption
is due to gas located in the lower halo of the Milky Way (see, e.g., 
Welty et al.\,1999), so that we can make use of the simple formalism
described by R03 to find the hydrogen volume density
($n_{\rm H}$) required to describe the observed H$_2$ column density 
in a formation-dissociation equilibrium:

\begin{equation}
n_{\rm H} \approx 9.2 \times 10^5 N({\rm H_2}) N({\rm H\,I})^{-1} S \phi^{-1}. 
\end{equation}

The parameters $S$ and $\phi$ represent scaling factors described
below.
We assume that the H$_2$ grain formation rate in the
IVC is similar to that in the disk of the Milky Way, and that the
H$_2$ photo-absorption rate at the edge of the IVC is 
half of that typically found
in local interstellar gas subject to the (expected) reduced UV photon flux in
the halo of the Milky Way. 
The parameter $\phi$ in the above equation is a
scaling factor that relates the H\,{\sc i} and H$_2$ volume densities
with their column densities. 
A detailed discussion about these parameters
is presented in R03. Note that the high ionization fraction (Sect.\,4.1)
possibly indicates that the overall conditions in the IVC may not
account for this simple formation-dissociation scenario; this
will be further discussed in Sect.\,4.4.
Some of the 
neutral and molecular hydrogen may reside in physically distinct regions,
and not all of the H\,{\sc i} is available for the neutral-to-molecular
hydrogen conversion. With $\phi$ we separate the cold neutral medium (CNM),
in which the H$_2$ resides, from the warm neutral medium (WNM) that 
surrounds the CNM in the IVC.
In practice, it is difficult to estimate $\phi$
since its value characterizes the physical structure of the IVC, which
is unknown. The difference in $b$ values found for the 
atomic species ($b=5.9^{+2.3}_{-3.4}$ km\,s$^{-1}$) and the molecular hydrogen
($b=1.5^{+0.8}_{-0.2}$ km\,s$^{-1}$) is a strong argument 
for assuming a pronounced core-envelope structure (CNM versus WNM). It is 
not clear, however, what fraction of the total IVC H\,{\sc i} column
density is related to the innermost dense core in which the
H$_2$ resides, and what fraction is due to absorption from the 
(much more extended) lower-density envelope.
Similar to our previous study (R03), we will assume
that $\phi = 0.5$.
Equation (1) also
includes an H$_2$ line self-shielding factor. The relatively high 
column density found for the H$_2$ (log $N($H$_2)=16.6 \pm 0.5$,
the low value for T$_{01}=51 \pm 11$ K,
and the very low $b$ value imply
that the H$_2$ absorption in the IVC towards Sk -68\,80 arises in
a dense, single-component cloudlet, for which H$_2$ line self-shielding 
has to be taken into account. We use the approximation provided by
Draine \& Bertoldi (1996), who find that for log $N($H$_2)>14$ the
UV dissociation rate in the cloud core is reduced by the factor 
$S=(N($H$_2)/10^{14}$cm$^{-2})^{-0.75}$ (we here use the letter $S$ instead of
their $f_{\rm shield}$ in order to avoid confusion 
with the molecular hydrogen fraction, $f$, defined in Sect.\,4.2).
In our case, we have 
$N($H$_2)=4.0 \times 10^{16}$ cm$^{-2}$,
$S\approx 0.011$, 
and we assume
that $N($H\,{\sc i})$=1 \times 10^{18}$ cm$^{-2}$, as indicated
by the O\,{\sc i} data. From equation (1) we then obtain
$n_{\rm H}=810$ cm$^{-3}$, and the linear diameter of the H$_2$ bearing
structure, $D=\phi N($H\,{\sc i}$) n_{\rm H}^{-1}$, is only 
$6.2 \times 10^{14}$ cm, or $\sim 41$ AU.
It thus appears that the molecular hydrogen in the IVC towards 
Sk -68\,80 is situated in a very small dense
filament, indicating the presence of substantial small-scale
structure in this diffuse halo cloud.

\subsection{Possible complications}

The results that we have obtained above are derived
by a straightforward analysis of the H$_2$ and metal
line absorption in the IVC component in front
of Sk -68\,80. 
The high density 
and the small size of the molecular 
structure, inferred from calculating
the H$_2$ abundance in a formation-dissociation 
equilibrium, are remarkable. In view of these
results, it is important to consider and discuss
possible complications and systematic errors 
that might have influenced 
our analysis, and to point to future 
observations that could help to confirm or 
discard the interpretations that we present 
in this paper.
In the following paragraphs, we list several possible 
complications:\\
\\
(1) {\it Significant velocity structure is 
present in the atomic IVC gas, but is unresolved in
the FUSE data.} The presence of velocity structure
in the IVC appears likely given the many
sub-components in the IVC gas towards SN 1987A
seen in 
very-high resolution optical spectra (Welty et al.\,1999).
Such unresolved velocity structure in the lower-resolution
FUSE data may introduce a significant uncertainty for
the determination of heavy element column densities, for which 
we had assumed a single Gaussian component (see Sect.\,4.1)
with a $b$ value of $5.9$ km\,s$^{-1}$. If several
sub-components with lower $b$-values are present, we
might underestimate the column density of O\,{\sc i} 
(and the other elements) and thus the total H\,{\sc i} gas
column density along the line of sight used for equation (1). Also, it is
possible that the various atomic species (Tables 1 \& 2) have different 
$b$ values
because of the ionization structure in the gas. This introduces
another uncertainty for the atomic column densities listed in
Table 2. High-resolution optical data for Sk -68\,80 will help
to investigate possible sub-component structure in the IVC gas.\\
\\
(2) {\it The metallicity of the gas is lower than solar.}
We had assumed a solar metallicity for estimating the
neutral hydrogen column density in the IVC towards 
Sk -68\,80, assuming that this IVC has abundances similar
to other IVCs in the Milky Way halo (e.g., Richter et al.\,2001c). 
If the actual metallicty of the gas is 
lower (for example, if the gas belongs to the LMC rather 
than to the Milky Way), we will underestimate $N$(H\,{\sc i})$_{\rm IVC}$
using this method, and the parameters $n_{\rm H}$ and $D$ 
derived from equation
(1) would have to be corrected. The H\,{\sc i} 21cm 
data gives no evidence that we have significantly
underestimated $N$(H\,{\sc i})$_{\rm IVC}$ by this method,
but radio beam smearing may complicate 
such a comparison. If the metallicity is lower than solar, the
dust abundance in the gas should be reduced as well, 
so that the H$_2$ grain formation rate in the IVC 
(see equation (2) in R03)
should be smaller than for solar-metallicity gas.
In this case, we would overestimate the H$_2$ formation
rate and underestimate the hydrogen volume 
density ($n_{\rm H}$) that is required to balance
the H$_2$ formation/dissociation at the observed column densities
(see equation (1)).\\
\\
(3) {\it The dissociating UV radiation field is lower or higher.}
We have estimated the dissociating UV radiation field in the
lower Milky Way halo (see R03) using the scaling relation 
provided by Wolfire et al.\,(1995). 
Assuming that the IVC is located $\sim 1$ kpc above the
Galactic plane, the UV radiation field is expected to be reduced
by a factor of $\sim 2$ in comparison to the midplane
intensity, mainly because of extinction by dust grains.
If the position of the IVC in the halo of the Milky Way is
such that the UV field at the IVC is 
much lower than assumed (e.g., due to shielding effects),
then the H$_2$ photo-dissociation rate, and thus $n_{\rm H}$,
would be overestimated. However, the high degree of ionization
(see Sect.\,4.1) may also imply that the UV field is much {\it stronger} than
assumed, leading to an enhanced photoionization of the IVC. If so, we would 
underestimate the H$_2$ photo-dissociation, and thus $n_{\rm H}$.\\
\\
(4) {\it The WNM dominates the neutral hydrogen column density.}
Yet another uncertainty is introduced 
through the factor $\phi$ which we have used in equation (1) to
account for the possibility that not all of the neutral material
is physically related to the CNM and the molecular gas. 
If the WNM is the main contributor to the IVC H\,{\sc i} column
density, $\phi$ could be much smaller than the assumed $\phi=0.5$.
In this case, 
$n_{\rm H}$ would be underestimated.\\
\\
(5) {\it The H$_2$ gas is not in formation-dissociation
equilibrium.} Equation (1) describes the hydrogen volume
density that is necessary to balance the formation of H$_2$
on dust grains with the dissociation by UV photons at 
a fractional abundance of H$_2$ (in terms of column density)
that is provided by the observations. However, such an equilibrium
situation might not be appropriate, in which case our conclusions
about the hydrogen volume density and diameter of the 
structure would be incorrect. Evidence for a possible
non-equilibrium situation is provided by the high
ionization fraction in the IVC gas, which may indicate
the presence of a shock that collisionally ionizes
the gas.\\
\\
(6) {\it The atomic IVC gas and the molecular hydrogen at 
$+50$ km\,s$^{-1}$ are not related.} We have assumed that the
IVC H$_2$ absorption towards Sk -68\,80 
is related to the widespread neutral and ionized material
at intermediate velocities in front of the LMC that is
seen along many sight lines (see, e.g., Danforth et al.\,2002).
It is possible, however, that the H$_2$ absorption occurs
in gas that is spatially and/or physically unrelated
to the neutral IVC gas, coincidentally having a similar
radial velocity. Theoretically, the H$_2$ absorption
at similar velocities could be somehow related to circumstellar
material or gas from supernova remnants (e.g., Welsh, Rachford,
\& Tumlinson 2002), in which case our conclusions 
may be incorrect. Dense molecular clumps in the outskirts
of our Galaxy have been proposed as candidates for
baryonic dark matter (e.g., de Paolis et al. 1995;
Pfenniger, Combes \& Martinet 1994). 
The H$_2$ absorption
at intermediate velocities may be due to diffuse inter-clump
gas that could arise from H$_2$ clump collisions in the halo, 
and that would be spatially much more extended than the dense 
clumps. Such gas probably would have a 
very low metal and dust content, and the parameters chosen for
equation (1) would be invalid.

\section{The HVC near $+120$ km\,s$^{-1}$}

The HVC component near $+120$ km\,s$^{-1}$ shows slightly
stronger atomic absorption than the IVC component (see Fig.\,2),
and we have analyzed the HVC absorption in a similar fashion
as for the IVC.

Equivalent widths and upper limits for C\,{\sc i}, O\,{\sc i},
Si\,{\sc ii}, P\,{\sc ii}, Ar\,{\sc i}, and Fe\,{\sc ii} are
listed in Table 1. The atomic data fit on a curve of growth
with $b=30.0^{+9.2}_{-5.4}$ km\,s$^{-1}$. This rather high $b$ value implies
the presence of unresolved sub-structure and/or substantial
turbulence within the gas. 
Logarithmic column densities for the species listed 
above, as derived
from the single-component curve of growth with $b=30.0$ km\,s$^{-1}$,
are presented in Table 2. Due to the probable existence of unresolved
sub-structure and the uncertain H\,{\sc i} column density (see Sect.\,3 
and Appendix), we do not derive gas-phase abundances for this cloud.
The relatively high Fe\,{\sc ii} and Si\,{\sc ii} column densities
in comparison to O\,{\sc i} ([Fe\,{\sc ii}/O\,{\sc i}]$=+0.6$ 
and [Si\,{\sc ii}/O\,{\sc i}]$=+0.3$) suggest a high degree of ionization,
similar to what is found for the IVC gas (see also Bluhm et al.\,2000).

Molecular hydrogen in the HVC is possibly detected in a few lines
for $J\geq 3$ (Fig.\,4 and Table 1), but the features are too weak to claim
a firm detection. However, the presence of HVC H$_2$ along the nearby
sight line towards Sk -68\,82 (see Appendix) may imply that these
features indeed are related to H$_2$ absorption in the HVC component.
Upper limits for the H$_2$ column densities in the HVC gas towards
Sk -68\,80 have been derived assuming $b\geq1.0$ km\,s$^{-1}$;
they are listed in Table 2.

\section{Discussion}

\begin{figure}
\resizebox{1.0\hsize}{!}{\includegraphics{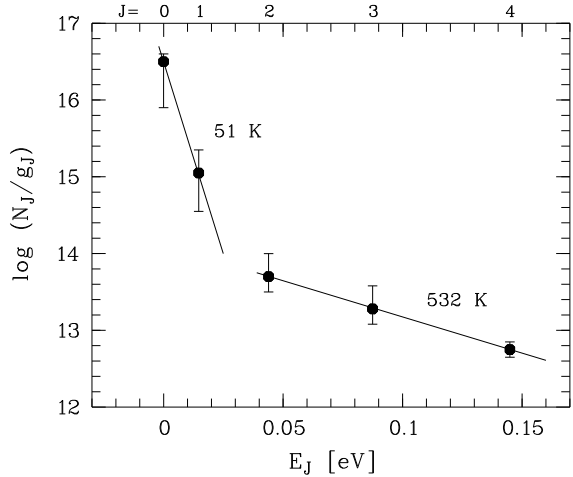}}
\caption[]{
Rotational excitation of the H$_2$ gas in the IVC
towards Sk -68\,80. The two lowest rotational states ($J=0,1$) can be
fitted to an equivalent Boltzmann temperature of 
T$_{01}=51\pm11$ K, while the higher rotational states
($J=2-4$) fit to a temperature of T$_{24}=532\pm124$ K. 
}
\end{figure}

The evidence for the existence of sub-pc structure
in the diffuse interstellar medium has been
accumulating impressively over the last few years, and 
is based on independent observations using various
different observation techniques, such as
H\,{\sc i} 21cm absorption lines studies (e.g., Faison et al.\,1998)
and optical absorption line studies (e.g., Lauroesch, Meyer, \&
Blades 2000). Observations of
diffuse molecular hydrogen, as shown in this study, 
may represent
yet another, independent method to study the nature
of the ISM at very small scales, assuming that the
parameters that we used for our H$_2$ formation-dissociation
equilibrium calculation are roughly correct. 
The hydrogen volume density derived in this study suggests
that the intermediate-velocity H$_2$ gas in front of
Sk -68\,80 may be related to the tiny-scale atomic
structures (TSAS, Heiles 1997) that have been found
in H\,{\sc i} 21cm absorption line studies. This is
also supported by the fact that
the H$_2$ excitation temperature of T$_{01}=51\pm11$ K 
corresponds to the canonical value of the 
CNM (Heiles 1997), in which the TSAS
are expected to be embedded. Recently, small scale structure
in the ISM has also been found in CO emission (Heithausen 2002).
It is possible that the H$_2$ gas detected here samples the 
transition zone from the cold neutral gas
to the dense molecular gas phase at 
small scales.

While more and more observations indicate that small-scale
structure represents an important aspect of the ISM,
very little is known about the overall physical properties.
At a temperature of T$\sim50$ K and a density of $n\sim10^3$ cm$^{-3}$
the thermal pressure, $P_{\rm TSAS}/k=n$T, is $\sim 5\times 10^4$ cm$^{-3}$,
about 13 times higher than the standard thermal pressure 
in the CNM. Although the turbulent pressure may dominate
the total gas pressure in TSAS, it remains unclear whether
it could account for this large discrepancy.
Heiles (1997) offers several geometrical solutions to
account generally for the pressure problem in the TSAS,
motivated by the exceptionally high volume
densities ($\sim 10^5$ cm$^{-3}$) inferred from VLBI observations.
He finds that if the TSAS are associated 
with curved filaments and sheets rather than with spherical
clouds one could bring the high ``apparent'' volume densities from
the H\,{\sc i} observations down to a level of $\sim 10^3$ cm$^{-3}$,
thus into the density range we have obtained by a 
completely different method. Still, our density
estimate from the formation-dissociation equilibrium
of H$_2$ is not independent of the geometry of the 
absorbing structure:
if the IVC H$_2$ absorption would occur
in a sheet or curved filament rather than in a spherical cloud, this
would change the geometry for the H$_2$ self-shielding.
For an elongated filament with an aspect ratio of
four we would overestimate the self-shielding and 
underestimate the actual volume density for the 
H$_2$ formation-dissociation equilibrium by a
factor of $\sim 3$. However, the fact that we
additional effort to account for the complex 
formation and dissociation processes of molecules in such filaments
with rather complex geometries.
It also remains unknown, whether these structures are 
related to even smaller and denser structures that may
contain a significant amount of baryonic (molecular)
dark matter (e.g., Pfenniger, Combes, \& Martinet 1994).

One interesting aspect of the detection presented in 
this paper concerns the line self-shielding of the H$_2$.
Since the efficiency of H$_2$ self-shielding mostly depends on
the H$_2$ column density,
small filaments with low neutral gas column densities 
(such as the IVC H$_2$ filament towards Sk -68\,80)
are not able to shield their molecular interior completely from the
dissociating UV radiation. This probably prevents the
the formation of CO (but see Heithausen (2002) for 
higher-column density gas)
and keeps the gas
from turning completely molecular. At a given
volume density distribution and H$_2$ grain
formation rate, the molecular
gas fraction at every point in
such a filament is determined completely by the
intensity of the ambient UV radiation field.
Thus, if the volume density distribution
in the filament does not change dramatically
in time,
the UV field stabilizes the molecular fraction
in the filament at a moderate level
and may prevent a
further fragmentation. 
Switching off the external UV field would
rapidly increase the molecular fraction at each point,
the self-shielding would become more efficient,
and the structure may turn completely molecular.
If the ISM favors the 
formation of low-column density 
filamentary structure instead of the large-column
density clouds, this could be a very efficient way
to suppress rapid star formation in dynamically
quiescent regions of galaxies
\footnote{Here, dynamically quiescent means
that no local star formation and supernova explosions
are present, which would dominate the evolution of the
surrounding ISM by way of shocks and compression.},
because the gas
is confined to very small gas pockets that
cannot turn completely molecular due to the
lack of efficient self-shielding.

The many detections
of H$_2$ in IVCs (R03) imply that halo clouds respresent
an excellent laboratory to study diffuse molecular gas
and its small-scale structure 
because of the velocity separation of these clouds from
strong local disk components and the moderate gas column
densities that characterize these clouds.
In the local ISM, such small-scale H$_2$ filaments 
(if they exist) might
be invisible because their radial velocities
along a given line of sight through the disk would not be 
significantly different from those of the high-column
density disk clouds. High-column density
absorbers would clearly dominate 
the H$_2$ absorption spectrum and completely overlap the much
weaker absorption caused by low-column density
filaments. Such filaments in the disk 
would therefore remain unnoticed. The detection of H$_2$ in
solar-metallicity IVCs in comparison to the non-detection of H$_2$ 
in the metal-poor HVC Complex C
(Richter et al.\,2001b) supports our original idea that 
observations of H$_2$ are helpful to distinguish 
between the various processes that are responsible
for the phenomenon of IVCs and HVCs in the Milky Way halo
(Richter et al.\,1999).

With the large number of UV bright stars distributed over
a relatively small area of the sky, the LMC provides 
an excellent backdrop to study small-scale structure
in the halo IVC and HVC gas in front of it. 
Additional high S/N FUSE data would be helpful in searching for other
directions in which H$_2$ halo absorption might be present. 
High-resolution optical data for Sk -68\,80
and other sight lines 
are required to better understand the velocity structure
of the halo gas in front of the LMC and to derive accurate $b$ values. This 
will be crucial to test the conclusions we have 
drawn in this paper from the intermediate-resolution FUSE data. 

\begin{acknowledgements}
      This work is based on data obtained for the
      the Guaranteed Time Team by the NASA-CNES-CSA FUSE
      mission operated by the Johns Hopkins University.
      Financial support has been provided by NASA
      contract NAS5-32985. 
      P.R. is supported by the
      \emph{Deut\-sche For\-schungs\-ge\-mein\-schaft}.
      JCH recognizes support from 
      NASA grant NAG5-12345.
      We thank K.S. de\,Boer and W.P. Blair for helpful comments.
\end{acknowledgements}

\appendix

\section{Sk -68\,82 and other LMC sight lines}

\subsection{LMC sight lines sampled by FUSE}

The LMC consists of a large number of UV
bright stars that are, in principle, suitable as
background sources for absorption-line spectroscopy
of intervening interstellar material. So far,
FUSE has observed several dozen stars in the
LMC as part of various Principle-Investigator (PI)
and Guest-Investigator (GI) programs.
An atlas of FUSE spectra of Magellanic Cloud stars
is provided by Danforth et al.\,(2002).
We have taken a closer look at the
FUSE LMC data to identify other sight lines
that could be used to study intermediate- and high-velocity
H$_2$ gas.
In many cases, absorption
from the IVC and HVC
components are quite weak (a good
indicator for this is the strong Fe\,{\sc ii} $\lambda 1144.938$
line; see Figs.\,3-59 from Danforth et al.\,2002). In other
spectra the S/N is low, or the stellar continuum
has a very irregular shape at small scales ($\leq 1$ \AA).
For these cases, the identification of H$_2$ at IVC and HVC
velocities is hampered by the low data quality.
Only a few sight lines (e.g., Sk -67\,101 and Sk -67\,104)
exhibit relatively strong IVC/HVC absorption at good S/N
and a reliable continuum,
but no convincing evidence for H$_2$ absorption
in the halo components is found from a first inspection.
These spectra, however,
will be useful to study in detail the atomic gas in the IVC and
HVC components in combination with high-resolution
optical data that will be necessary to disentangle the
sub-component structure.

\subsection{Sk -68\,82}

\begin{figure}[t!]
\resizebox{1.3\hsize}{!}{\includegraphics{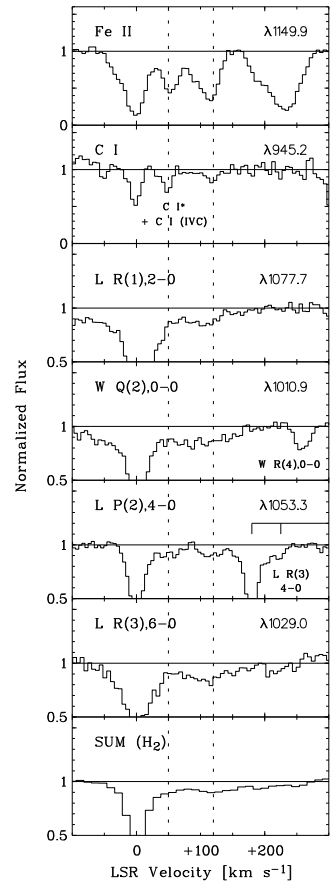}}
\caption[]{
Atomic and H$_2$ absorption profiles in direction of 
Sk -68\,82. The strong Fe\,{\sc ii} line (upper panel)
suggests a very similar component strucutre as for 
Sk -68\,80. In contrast to Sk -68\,80 (see Fig.\,2), 
C\,{\sc i} absorption is present in the HVC component.
Weak molecular hydrogen absorption is present at IVC and HVC
velocities, but the individual H$_2$ line profiles have very irregular
shapes due to small-scale structure in the continuum flux
at wavelengths between $1000$ and $1080$ \AA. A cumulative
H$_2$ profile from a co-addition of 15 lines is
presented in the lower-most panel.
}
\end{figure}

One special case that we want to highlight
is the spectrum of Sk -68\,82 (HD\,269546), the
sight line where the phenomenon of H$_2$ absorption
in intermediate- and high-velocity gas was found
for the first time in
low S/N ORFEUS data (Richter et al.\,1999; Bluhm et al.\,2000).
The ORFEUS H$_2$ findings in the IVC/HVC gas in front
of the LMC were coincidental detections during a project
searching for H$_2$ absorption in the LMC (Richter 2000).
The presence of IVC/HVC H$_2$
is evident at a $6 \sigma$ level
in the composite
velocity profile of H$_2$ for which we had co-added
various H$_2$ transitions
for $J\leq6$ to study the general velocity distribution
of the H$_2$ towards Sk -68\,82 in the ORFEUS data. In individual lines,
however, H$_2$ is detected at low significance ($1-4 \sigma$;
see Richter et al.\,1999, Bluhm et al.\,2000) due to the
low S/N in the data, so that the H$_2$ column densities,
$b$ values, and excitation temperatures derived for the
IVC and HVC gas are quite uncertain.

We have re-investigated this sight line with much higher
quality FUSE data of Sk -68\,82 to check the previous
results and conclusions. The FUSE data for Sk -68\,82
(program IDs P2030101-P2030104) 
were reduced with the {\tt CALFUSE} v2.05
pipeline in a fashion similar to the data for Sk -68\,80.
A detailed inspection of the spectrum shows
that the continuum flux of Sk -68\,82
is much more irregular and complicated than for
Sk -68\,60, in particular at scales $\leq 1$\AA.
These irregularities in the continuum complicate
the interpretation of interstellar absorption
more than was evident from the lower quality
ORFEUS data.
Fig.\,A.1 shows several atomic and H$_2$ absorption profiles
for Sk -68\,82. For each line we show
the normalized flux plotted against the LSR velocity.
The profiles are normalized to a smooth continuum
that describes the background flux on scales 
$\geq 1$ \AA. We cannot account for the many
structures and features in the continuum 
at smaller scales ($\leq1$ \AA), so that
absorption components with small equivalent
widths (such as the IVC and HVC H$_2$ absorption)
exhibit quite irregulary shaped absorption profiles.
The two upper panels show absorption
by Fe\,{\sc ii} $\lambda1144.989$ and
C\,{\sc i} $\lambda 945.151$. These two
lines lie in regions of the spectrum
where the choice of the continuum is
less critical than for the regions
in which most of the H$_2$
lines are located ($1000\leq
\lambda \leq 1080$\AA). The velocity
distribution of Fe\,{\sc ii} absorption
is very similar to that of Sk -68\,80
(see Fig.\,2). C\,{\sc i} absorption
towards Sk -68\,82 is seen not only
in the local Galactic gas, but also in
the HVC component near $+120$
km\,s$^{-1}$. The IVC component is
(as for Sk -68\,80) blended by local
C\,{\sc i}* absorption. The presence of
C\,{\sc i} at $+120$ km\,s$^{-1}$
suggests the presence of a cool, dense gas
component in the HVC, since C\,{\sc i}
is easily ionized in warm diffuse gas
(the ionization potential of C\,{\sc i} is $11.3$ eV).
The atomic HVC gas
towards Sk -68\,82 has a much lower $b$ value
($\sim 12$ km\,s$^{-1}$), but higher column densities
($\sim 0.7$ dex for
O\,{\sc i} and $\sim 0.3$ dex for Fe\,{\sc ii}
and Si\,{\sc ii}) than the HVC component towards
Sk -68\,80. Obviously, small scale structure
exists on scales that separate these two stars
on the sky ($\sim 1$ arcmin). Thus, the new FUSE data
imply that the 32 arcmin beam H\,{\sc I} 21cm
column density is, despite earlier
attempts (Richter et al.\,1999; Bluhm et al.\,2000),
not a good reference to 
calculate precise gas-phase abundances for this cloud.

The interpretation of the H$_2$ absorption
towards Sk -68\,82 is much more difficult
than for Sk -68\,80 due to the difficult
continuum situation in the wavelength range
of the H$_2$ Lyman- and Werner bands.
Fig.\,A.1 shows some examples for the H$_2$ absorption
line profiles towards Sk -68\,82.
For many lines, in particular for rotational states
$J\geq2$, H$_2$ absorption extends from $-30$ to $+150$ km\,s$^{-1}$,
but is overlapped by the small-scale structure in the continuum.
In order to minimize the effects of the (randomly
distributed) continuum small-scale structure,
we have co-added $15$ H$_2$ lines from the rotational
states $J=1-3$ (Fig.\,A.1, lower-most panel) to analyze
the general H$_2$ velocity distribution in the FUSE data of
Sk -68\,82. As the cumulative H$_2$ absorption profile
confirms, H$_2$ is present at IVC and HVC velocities,
but is smeared over the velocity range from $-20$ to $+140$
km\,s$^{-1}$, and a clear component structure is still not readily
visible. 
There are significant discrepancies in the shape of some H$_2$ lines 
between the FUSE and the older ORFEUS data (e.g., W Q(2),0-0 
$\lambda 1010.9$; see Richter et al.\,1999 and Fig.\,A.1). 
Given the low S/N in the ORFEUS data and the resulting $1\sigma$
uncertainties (Table 1 in Richter et al.\,1999), these differences can be 
easily explained by noise structures in the ORFEUS data.
However, since the background star (together with the LMC) 
has a substantial transversal motion behind the 
Milky Way halo gas, 
such differences could also arise from
small-scale structure within the HVC H$_2$ gas, considering the
results for the IVC H$_2$ gas towards Sk -68\,80 and
the fact that the ORFEUS data for 
Sk -68\,82 was taken $\sim4$ years before the FUSE data.
Temporal variations of absorption lines in diffuse interstellar gas
have been reported by Lauroesch, Meyer, \& Blades (2000). Unfortunately,
the S/N in the ORFEUS data is too low to test this
interesting idea, but future FUSE observations 
will help to search for such temporal variations.
The absorption depths for the IVC and HVC H$_2$ absorption
in the FUSE data of Sk -68\,82
correspond to total H$_2$ column densities of log $N$(H$_2)=14-16$,
depending on the adopted $b$ value.
We have to realize at this point that we are
unable to improve our knowledge about the molecular
material in the IVC and HVC towards Sk -68\,82 with
the high S/N FUSE data.
To correct our previous results
(Richter et al.\,1999, Bluhm et al.\,2000)
from ORFEUS to a more conservative statement we can now state that
H$_2$ is present
in the IVC and HVC towards Sk -68\,82, but the high
S/N FUSE data show that a determination of precise
column densities is impossible due to small-scale
structure in the continuum. Similarly, without having
reliable values for $N(J)$, we cannot derive
accurate excitation temperatures for the IVC and HVC gas.
The fact that H$_2$ at IVC and HVC velocities is
seen in levels up to $J=4$, however, implies a relatively high
degree of rotational excitation, as was already concluded
from the ORFEUS data (Richter et al.\,1999; Bluhm et al.\,2000).
The two stars Sk -68\,80 and Sk -68\,82 are separated by only
$\sim 1$ arcmin. The presence of H$_2$ in the IVC towards
both stars suggests that the IVC gas in this general direction
of N\,144 in the LMC consists of dense, cool material from
which H$_2$ bearing filaments can form. A similar
conclusion holds also for the HVC component, in which
H$_2$ is present towards Sk -68\,82 and possibly also
towards Sk -68\,80 (see Sect.\,4). 

\end{document}